\documentclass[11pt]{article}
\usepackage[top = 2 cm, bottom = 2.5 cm, left = 2.5 cm, right = 2 cm]{geometry}
\usepackage{authblk, cite, color, amssymb, amsmath, enumitem}
\usepackage[hypertex, colorlinks = true, linkcolor = blue, citecolor = red]{hyperref}
\usepackage[dvipsnames]{xcolor}

\begin{document}

\title{Split octonionic Dirac equation}

\author[1,2]{Merab Gogberashvili \thanks{gogber@gmail.com}}
\author[1,3] {Alexandre Gurchumelia \thanks{alexandre.gurchumelia@gmail.com}}

\affil[1]{Javakhishvili State University, 3 Chavchavadze Ave., Tbilisi 0179, Georgia}
\affil[2]{Andronikashvili Institute of Physics, 6 Tamarashvili St., Tbilisi 0177, Georgia}
\affil[3]{Kharadze Georgian National Astrophysical Observatory, Abastumani 0301, Georgia}

\maketitle

\begin{abstract}
\noindent
The novel forms of the split octonionic Dirac equation and its corresponding Lagrangian are derived using symbolic computing techniques.

\end{abstract}

\vskip 1cm

The Dirac equation is a cornerstone of theoretical physics, providing a framework for understanding the behavior of fermions within the realm of relativistic quantum mechanics. It has been extensively studied in various contexts and has provided a powerful framework for understanding the properties of elementary particles. Expanding upon its traditional formulations, recent research has explored novel mathematical structures to deepen our comprehension of fundamental physics.

In this study, we delve into the Dirac equation within the framework of the non-associative algebra of split octonions \cite{Sc, Sp-Ve, Baez}. The quatratic form on split octonions has pseudo-Euclidean structure and may provide new insights into the intrinsic nature of spacetime \cite{Gog-1, Gog-2, Go-Sa, Go-Gu} and formulations of relativistic equations \cite{Gogberashvili:2005cp, Gogberashvili:2005xb, Chanyal:2015sja, Chanyal:2010sz}. Since for the 8-dimensional non-associative algebra of split octonions calculations are cumbersome, we have developed a Python module called SplitOct, leveraging SymPy for symbolic computations \cite{Meurer:2017yhf}. Source code for SplitOct and computation examples can be found in the repository \cite{Gurchumelia2023}.

In the standard formulation, the Lagrangian for a Dirac field $\psi$ is given by:
\begin{equation} \label{eq:matrix_Lagrangian}
\mathcal{L} = \overline{\psi}\left(i\gamma^{\mu}\partial_{\mu}-m\right)\psi + \mathrm{h.c.}~, \qquad \qquad (\mu = 0, 1, 2, 3)
\end{equation}
where $\overline{\psi} = \psi^{\dagger}\gamma^{0}$ and $\partial_{\mu} = \left(\partial_{t},\nabla\right)$. For the Dirac gamma matrices in (\ref{eq:matrix_Lagrangian}) we use the representation,
\begin{equation}
\gamma^{0} =
\begin{pmatrix}
1_{2\times2} & 0\\
0 & -1_{2\times2}
\end{pmatrix},\qquad
\gamma^{n} =
\begin{pmatrix}0 & \sigma_{n}\\
-\sigma_{n} & 0
\end{pmatrix},
\qquad \qquad (n = 1, 2, 3)
\end{equation}
where $\sigma_{n}$ represents the standard Pauli matrices.

Dirac field in an external four-potential $A_{\mu} = \left(\phi,- A_n\right)$ is described by the equation
\begin{equation} \label{Dirac}
i\gamma^{\mu}D_{\mu}\psi = m\psi~,
\end{equation}
where $D_{\mu}=\partial_{\mu}+ieA_{\mu}$. If we label real and imaginary components of the bispinorial wavefunction as
\begin{equation}
\psi = \left(
\begin{aligned}
\psi_{4} & + i\psi_{7}\\
-\psi_{6} & + i\psi_{5}\\
\psi_{3} & + i\psi_{0}\\
\psi_{1} & + i\psi_{2}
\end{aligned}
\right),
\end{equation}
the equation (\ref{Dirac}) can be write as a following system:
\begin{equation} \label{eq:dirac_components}
\begin{aligned}
-\phi\psi_{4}+A_{1}\psi_{1}+A_{2}\psi_{2}+A_{3}\psi_{3}-m\psi_{4}-\partial_{t}\psi_{7}-\partial_{x}\psi_{2}+\partial_{y}\psi_{1}-\partial_{z}\psi_{0} & =0~,\\
\phi\psi_{6}+A_{1}\psi_{3}-A_{2}\psi_{0}-A_{3}\psi_{1}+m\psi_{6}-\partial_{t}\psi_{5}-\partial_{x}\psi_{0}-\partial_{y}\psi_{3}+\partial_{z}\psi_{2} & =0~,\\
\phi\psi_{3}+A_{1}\psi_{6}-A_{2}\psi_{5}-A_{3}\psi_{4}-m\psi_{3}+\partial_{t}\psi_{0}+\partial_{x}\psi_{5}+\partial_{y}\psi_{6}+\partial_{z}\psi_{7} & =0~,\\
\phi\psi_{1}-A_{1}\psi_{4}+A_{2}\psi_{7}-A_{3}\psi_{6}-m\psi_{1}+\partial_{t}\psi_{2}+\partial_{x}\psi_{7}+\partial_{y}\psi_{4}-\partial_{z}\psi_{5} & =0~,\\
-\phi\psi_{7}+A_{1}\psi_{2}-A_{2}\psi_{1}+A_{3}\psi_{0}-m\psi_{7}+\partial_{t}\psi_{4}+\partial_{x}\psi_{1}+\partial_{y}\psi_{2}+\partial_{z}\psi_{3} & =0~,\\
-\phi\psi_{5}+A_{1}\psi_{0}+A_{2}\psi_{3}-A_{3}\psi_{2}-m\psi_{5}-\partial_{t}\psi_{6}+\partial_{x}\psi_{3}-\partial_{y}\psi_{0}-\partial_{z}\psi_{1} & =0~,\\
\phi\psi_{0}-A_{1}\psi_{5}-A_{2}\psi_{6}-A_{3}\psi_{7}-m\psi_{0}-\partial_{t}\psi_{3}+\partial_{x}\psi_{6}-\partial_{y}\psi_{5}-\partial_{z}\psi_{4} & =0~,\\
\phi\psi_{2}-A_{1}\psi_{7}-A_{2}\psi_{4}+A_{3}\psi_{5}-m\psi_{2}-\partial_{t}\psi_{1}-\partial_{x}\psi_{4}+\partial_{y}\psi_{7}-\partial_{z}\psi_{6} & =0~.
\end{aligned}
\end{equation}
We want to rewrite this system and corresponding Lagrangian (\ref{eq:matrix_Lagrangian}) using the algebra of split octonions.

Split octonions form an 8-dimensional non-associative algebra $\mathbb{O}^{\prime}$, which can be defined through the relations for their seven imaginary units:
\begin{equation} \label{algebra}
\begin{split}
& I^{2}=1~,\qquad j_{n}I=J_{n}~,\qquad j_{m}j_{n}=-\delta_{mn}+\sum_{\ell}\epsilon_{\ell mn}j_{\ell}~,\\
& J_{m}J_{n}=\delta_{mn}+\sum_{\ell}\epsilon_{\ell mn}j_{\ell}~,\quad J_{m}j_{n}=\delta_{mn}I-\sum_{\ell}\epsilon_{\ell mn}J_{\ell}~.\qquad\qquad\left(\ell,m,n,=1,2,3\right)
\end{split}
\end{equation}

General split octonionic number $x\in\mathbb{O}^{\prime}$ and its conjugate $\overline{x}\in\mathbb{O}^{\prime}$ can be written in the form:
\begin{equation}
\begin{split}
x & =x_{0}+Ix_{4}+\sum_{n=1}^{3}\left(j_{n}x_{n}+J_{n}x_{4+n}\right)~,\\
\overline{x} & =x_{0}-Ix_{4}-\sum_{n=1}^{3}\left(j_{n}x_{n}+J_{n}x_{4+n}\right)~,
\end{split}
\end{equation}
where $x_{0},x_{1},\ldots,x_{7}\in\mathbb{R}$. We use a quadratic form on split octonions $\mathcal{Q}:\mathbb{O}^{\prime}\rightarrow\mathbb{R}$,
\begin{equation}
\mathcal{Q}\left(x\right) = \overline{x}x ~,
\end{equation}
to introduce a non-degenerate bilinear form $\left\langle \cdot,\cdot\right\rangle :\mathbb{O}^{\prime}\times\mathbb{O}^{\prime}\rightarrow\mathbb{R}$, which is defined as
\begin{equation}
\begin{aligned}
\left\langle x,y\right\rangle  & = \frac{1}{2}\mathcal{Q}\left(x+y\right) - \frac{1}{2}\mathcal{Q}\left(x\right) - \frac{1}{2}\mathcal{Q}\left(y\right)=\\
 & = \frac{1}{2} \left(\overline{x}y + \overline{y}x\right) = \sum_{n=0}^{3}\left(x_{n}y_{n} - x_{4+n}y_{4+n}\right)~.
\end{aligned}
\end{equation}
For split octonionic functions $f:\mathbb{O}^{\prime}\rightarrow\mathbb{O}^{\prime}$ we can define differential operators:
\begin{equation} \label{partial}
\begin{split}
\partial & =\frac{1}{2}\left(\partial_{0}+I\partial_{4}\right)+\frac{1}{2}\sum_{n=1}^{3}\left(j_{n}\partial_{n}+J_{n}\partial_{4+n}\right)~,\\
\overline{\partial} & =\frac{1}{2}\left(\partial_{0}-I\partial_{4}\right)-\frac{1}{2}\sum_{n=1}^{3}\left(j_{n}\partial_{n}+J_{n}\partial_{4+n}\right)~,
\end{split}
\end{equation}
where $\partial_{n}$ is a partial derivative operator with respect to $x_{n}$ and the factor of a half ensures that
\begin{equation}
\partial x = \overline{\partial}\overline{x} = 1~, \qquad \overline{\partial}x = \partial\overline{x} = 0~.
\end{equation}
Note that for the definition (\ref{partial}), the expected rules of derivatives for high order polynomials do not work, namely $\partial x^{k} = kx^{k-1}$ does not hold. For non-split octonions this kind of differential operator and corresponding analyticity equation are studied in \cite{KauhanenOrelma_OctonionicAnalysis}.

Now let us obtain the Dirac equation for the split octonionic wavefunction, $\psi:\mathbb{O}^{\prime}\rightarrow\mathbb{O}^{\prime}$, in the external 4-potential vector field,
\begin{equation}
A = I\phi + \sum_{n=1}^{3}j_{n}A_{n}~.
\end{equation}
Consider a differential operator,
\begin{equation} \label{D}
D = 2I\partial I = \partial_{0} + I\partial_{4} - \sum_{n=1}^{3}\left(j_{n}\partial_{n}+J_{n}\partial_{4+n}\right)~,
\end{equation}
where $\partial$ is defined in (\ref{partial}), and assume that the octonionic wavefunction,
\begin{equation}
\psi = \psi_{0} + I\psi_{4} + \sum_{n=1}^{3}\left(j_{n}\psi_{n} + J_{n}\psi_{4+n}\right)~,
\end{equation}
is constant in variables $x_{0}, x_{5}, x_{6}$ and $x_{7}$. This assumption is equivalent to replacing the differential operator (\ref{D}) with:
\begin{equation}
D \rightarrow \mathcal{D} = I\partial_{t} - \sum_{n=1}^{3}j_{n}\partial_{n}~,
\end{equation}
where we made the replacement $\partial_{4} \to \partial_{t}$. Then, with the help of Python module SplitOct \cite{Gurchumelia2023}, it can be shown that split octonionic Dirac equation,
\begin{equation} \label{Dirac-oct}
\left(\mathcal{D} - J_{3}m\right)\psi = 0~,
\end{equation}
is component-wise equivalent to the standard Dirac system (\ref{eq:dirac_components}). We can also write the Lagrangian of free Dirac field in terms of split octonions:
\begin{equation} \label{Lagrangian}
\mathcal{L} = \frac{1}{2}\left\langle J_{3}\psi,\mathcal{D}\psi\right\rangle + \frac{1}{2}m\left\langle \psi,\psi\right\rangle ~.
\end{equation}

In conclusion, this paper has successfully derived the standard Dirac equation within the framework of split octonions. For calculations the Python module SplitOct software was used \cite{Gurchumelia2023}. Methods of calculations and obtained split octonionic Dirac equation (\ref{Dirac-oct}) and its Lagrangian (\ref{Lagrangian}) are slightly different from earlier findings in \cite{Gogberashvili:2005cp}. The main difference is the appearance of split octonionic imaginary unit $J_3$ in these expressions. This alteration has the potential to impact the final form of the split octonionic Standard Model Lagrangian, which remains an area for future exploration. This study opens up new avenues for research, paving the way for deeper insights into the fundamental structure of the universe.


\section*{Availability of data and materials}
The source code for SplitOct Python module used during the current study along with calculations in Jupyter notebooks are publicly available at: https://github.com/EQUINOX24/SplitOct



\begin{thebibliography}{12}

\bibitem{Sc} R.~Schafer,
\textit{Introduction to Non-Associative Algebras}
(Dover, NY 1995).

\bibitem{Sp-Ve} T.~A.~Springer and F.~D.~Veldkamp,
``Exceptional groups,''
in \textit{Octonions, Jordan Algebras and Exceptional Groups}, Springer Monographs in Mathematics
(Springer, Berlin 2000)
doi: 10.1007/978-3-662-12622-6\_7.

\bibitem{Baez} J.~C.~Baez,
``The octonions,''
Bull. Am. Math. Soc. \textbf{39} (2002) 145 [erratum: Bull. Am. Math. Soc. \textbf{42} (2005) 213],
doi: 10.1090/S0273-0979-01-00934-X
[arXiv: math/0105155 [math.RA]].

\bibitem{Gog-1} M.~Gogberashvili,
``Octonionic geometry,''
Adv. Appl. Clifford Algebras \textbf{15} (2005) 55,
doi: 10.1007/s00006-005-0003-2
[arXiv: hep-th/0409173].

\bibitem{Gog-2} M.~Gogberashvili,
``Rotations in the space of the split octonions,''
Adv. Math. Phys. \textbf{2009} (2009) 483079,
doi: 10.1155/2009/483079
[arXiv: 0808.2496 [math-ph]].

\bibitem{Go-Sa} M.~Gogberashvili and O.~Sakhelashvili,
``Geometrical applications of the split octonions,''
Adv. Math. Phys. \textbf{2015} (2015) 196708,
doi: 10.1155/2015/196708
[arXiv: 1506.01012 [math-ph]].

\bibitem{Go-Gu} M.~Gogberashvili and A.~Gurchumelia,
``Geometry of the non-compact G(2),''
J. Geom. Phys. \textbf{144} (2019) 308,
doi: 10.1016/j.geomphys.2019.06.015
[arXiv: 1903.04888 [physics.gen-ph]].

\bibitem{Gogberashvili:2005cp} M.~Gogberashvili,
``Octonionic version of Dirac equations,''
Int. J. Mod. Phys. A \textbf{21} (2006) 3513,
doi: 10.1142/S0217751X06028436
[arXiv: hep-th/0505101 [hep-th]].

\bibitem{Gogberashvili:2005xb} M.~Gogberashvili,
``Octonionic electrodynamics,''
J. Phys. A \textbf{39} (2006) 7099,
doi: 10.1088/0305-4470/39/22/020
[arXiv: hep-th/0512258 [hep-th]].

\bibitem{Chanyal:2015sja} B.~C.~Chanyal,
``Split octonion reformulation of generalized linear gravitational field equations,''
J. Math. Phys. \textbf{56} (2015) 051702,
doi: 10.1063/1.4921063.

\bibitem{Chanyal:2010sz} B.~C.~Chanyal, P.~S.~Bisht and O.~P.~S.~Negi,
``Generalized split-octonion electrodynamics,''
Int. J. Theor. Phys. \textbf{50} (2011) 1919,
doi: 10.1007/s10773-011-0706-1
[arXiv: 1011.3922 [physics.gen-ph]].

\bibitem{Meurer:2017yhf} A.~Meurer, \textit{et al.}
``SymPy: symbolic computing in Python,''
PeerJ Comput. Sci. \textbf{3} (2017) e103,
doi: 10.7717/peerj-cs.103.

\bibitem{Gurchumelia2023} A.~Gurchumelia,
``SplitOct'' (Computer software),
\texttt{https://github.com/EQUINOX24/SplitOct}.

\bibitem{KauhanenOrelma_OctonionicAnalysis} J.~Kauhanen and H.~Orelma,
``Cauchy-Riemann operators in octonionic analysis,''
Adv. Appl. Cliff. Alg. \textbf{28} (2018) 1,
doi: 10.1007/s00006-018-0826-2
[arXiv: 1701.08698 [math.CV]].

\end{thebibliography}
\end{document}